\documentclass[aps,prl,twocolumn,superscriptaddress,final]{revtex4-2}
\usepackage{amsmath}
\usepackage{amsfonts}
\usepackage{amssymb}
\usepackage{bbm}
\usepackage{graphicx}
\usepackage{xcolor}
\usepackage[colorlinks=True]{hyperref}
\usepackage{hypcap}
\usepackage{braket}
\usepackage{bm}
\usepackage{dsfont}
\usepackage{mathtools}
\usepackage[export]{adjustbox} 
\usepackage{verbatim}
\usepackage{amsthm}

\usepackage{csquotes}
\usepackage[backgroundcolor=white,bordercolor=red,linecolor=red,textcolor=red,textsize=footnotesize]{todonotes}

\newcommand{\xx}{z}

\newcommand{\lowerdist}{\lfloor x_{AB}/2\rfloor + 1}

\graphicspath{{.}{./tikz/}}

\DeclareMathOperator{\Tr}{Tr}

\newtheorem{theorem}{Theorem}
\newtheorem{lemma}{Lemma}

\newcommand{\dif}{{\rm d}}
\newcommand{\iu}{{\rm i}}

\begin{document}
	\def\papertitle{Lower bounds on the complexity of preparing mixed states}
	\title{\papertitle}
	\author{Max McGinley}
	\affiliation{TCM Group, Cavendish Laboratory, University of Cambridge, Cambridge CB3 0HE, UK}
	\author{Samuel J.~Garratt}
	\affiliation{Department of Physics, Princeton University, NJ 08544, USA}
	\date{\today}
	
	\begin{abstract}
		We establish a relationship between the correlations in a many-qubit mixed state and the minimum circuit depth needed for its preparation.
		If the mutual information between two subsystems exceeds the mutual information between one of those subsystems and the environment, which purifies the mixed state of the system, then the past lightcones of the subsystems must intersect one another. This results in a lower bound on the circuit depth of any ensemble of geometrically local unitaries that prepares the state to some specified degree of approximation. As an application, we derive lower bounds on the circuit depth needed to prepare thermal states of one-dimensional quantum critical systems described by conformal field theory, showing that the depth diverges as temperature is decreased up to a cutoff set by the preparation error. 
	\end{abstract}
	
	\maketitle

	\textit{Introduction.---}
	How quickly can a given quantum state be prepared on a quantum computer?
	By bringing this computational question into the domain of physics, new light has been shed on the nature of quantum phases of matter \cite{Bravyi2006,Chen2010}, the dynamics of isolated quantum systems \cite{cotler2017chaos, Roberts2017, Camargo2019, Ali2020}, and the holographic correspondence \cite{Susskind2016, Brown2016}. These insights point to a deep relationship between the physical properties of a state and its \textit{complexity}, i.e.~the minimum time required to prepare it within some specified model of quantum computation.
	
	Quantitative lower bounds on the complexity of pure states have been derived in a number of settings \cite{cotler2017chaos,Roberts2017,Aharonov2018, Brandao2021, Iaconis2021}, however analogous results for mixed states are less common \cite{Hastings2011, Eldar2021, Zhou2025}. This stems partly from the fact that there are many different ways to realise mixed states as ensembles or marginals of pure states \cite{Aharonov1998,Agon2019}. 
	Overcoming this limitation is important because many states that arise in nature are mixed---most notably thermal (Gibbs) states, whose preparation is essential for the quantum simulation of matter.
	
	In this work, we introduce a new way to construct lower bounds on the complexity of a given mixed state $\rho$ in terms of its correlations. Theorem \ref{thm:Exact} constrains the depth of any statistical ensemble of geometrically local circuits that prepares $\rho$, starting from a product initial state. Theorem \ref{thm:Approx} generalizes this result to the case of approximate state preparation. A technical challenge in relating complexity to correlations is that, unlike pure states, mixed states can exhibit long-ranged classical correlations that can be generated by local operations. We show how these classical correlations can be factored out by analysing purifications of $\rho$. Specifically, we show that if there are two degrees of freedom separated by a distance $x$ that are more correlated with one another than either is with the purifying system, then the required circuit depth is greater than $x/2$.

	These lower bounds on mixed state complexity are complementary to algorithms for preparing various mixed quantum states, which imply corresponding upper bounds \cite{Bakshi2025,rouze2024optimal,smid2025polynomial}. 
	A great deal of progress in this direction has been made for Gibbs states of local Hamiltonians \cite{Bakshi2025,rouze2024optimal,smid2025polynomial,brandao2019finite,Kuwuhara2020,Kuwuhara2021,Chen2023a,Chen2023b,rakovszky2024bottlenecks, Gilyen2024,Ding2025,Hahn2025,lloyd2025quantum}, including the discovery that high-temperature states can be prepared in zero depth \cite{Bakshi2025}. As a first application of our method, we study Gibbs states of one-dimensional systems, focusing on Hamiltonians at and near a quantum critical point described by conformal field theory (CFT). Since our bounds depend only on correlations in the Gibbs state, which can be computed using CFT, we can show that the required circuit depth diverges as temperature is decreased, up to a cutoff determined by the preparation error. 
	We conclude by discussing possible generalisations to state preparation schemes that use measurement-feedback loops, and connections to the study of mixed state topological order.

	\textit{Pure state preparation.---}We consider a system composed of $n$ qubits that live at the vertices of a graph $G$, and begin by reviewing the problem of preparing a pure state $\ket{\Psi}$. 
	A natural and well-studied way to characterise the complexity of $\ket{\Psi}$ is to ask whether there exists a unitary circuit $U$ of depth $d$ such that $\ket{\Psi} =  U\ket{0^{\otimes n}}$ (we consider exact preparation for now, but approximate preparation will be treated later on). Here, and for the rest of this work, we are considering geometrically local circuits, i.e.~those made up of two-qubit gates that act on pairs of qubits that share an edge in $G$.
	The minimum depth for which there exists such a circuit, which we denote as $d_{\rm min}(\Psi)$, quantifies the shortest possible time that it would take to prepare $\ket{\Psi}$ on a digital quantum computer with connectivity graph $G$. This serves as a measure of the complexity of $\ket{\Psi}$.

\begin{figure}
		\includegraphics[width=\columnwidth]{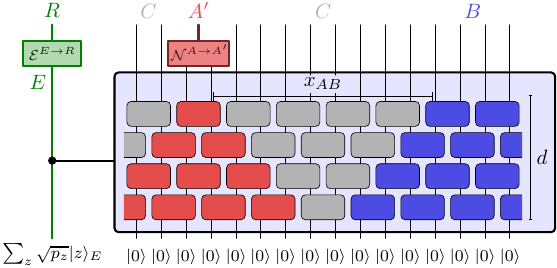}
		\caption{Starting from the initial state $\ket{0^{\otimes n}}$, a mixed state $\rho$ can be generated by applying randomly choosing unitaries $U_\xx$ with probabilities $p_\xx$. From such a preparation scheme, a purification can be constructed [Eq.~\eqref{eq:PurificationDef}], which can be viewed as the result of applying a $\xx$-controlled unitary $V = \sum_\xx \ket{\xx}\bra{\xx}_E \otimes U_\xx$, with the control $E$ initialized in the state $\sum_\xx \sqrt{p_\xx}\ket{\xx}_E$, and the system as the target. Dividing the system qubits into regions $A$,$B$,$C$, we denote $x_{AB}$ as the minimum graph distance between $A$ and $B$ (here a one-dimensional chain is shown). If each $U_\xx$ is a circuit of depth $d < \lowerdist$, then the past lightcones of $A$ and $B$ (red and blue shaded gates, respectively) do not overlap, and $I(A:B)_{\phi_\xx} = 0$. The classical-quantum state $\rho^{ABCR}_{\rm opt}$ [Eq.~\eqref{eq:RhoCQ}] can be generated by applying a dephasing channel $\mathcal{E}^{E \rightarrow R}$ to $E$.
		}
		\label{fig:Lightcone}
	\end{figure}

	Using the fact that correlations can only spread at a finite speed in  geometrically local circuits, a simple lower bound can be obtained. If we partition the graph of qubits $G$ into three regions $A, B, C$, where $A$ and $B$ are separated by a distance $x_{AB}$ (see Fig.~\ref{fig:Lightcone}), then any nonzero connected correlations $\braket{O_AO_B}_{c, \Psi} \coloneqq \braket{O_AO_B}_{\Psi} - \braket{O_A}_{\Psi}\braket{O_B}_{\Psi}$ between observables $O_{A,B}$ that are supported on $A,B$ imply that $\ket{\Psi}$ cannot be prepared exactly by a circuit of depth less than $x_{AB}/2$. This can also be expressed in an observable-independent way in terms of the mutual information
	\begin{align}
		I(A:B)_{\Psi} &> 0 
		&\Rightarrow d_{\rm min}(\Psi) \geq \lfloor x_{AB}/2\rfloor + 1
		\label{eq:MIPure}
	\end{align}
	Here $S(A)_{\Psi} = -\Tr[\rho^A\ln \rho^A]$ is the von Neumann entropy of the reduced state $\rho^A = \Tr_{A^c}[\Psi]$, and $I(A:B) \coloneqq S(A) + S(B) - S(AB)$. We use the shorthand $\Phi \equiv \ket{\Phi}\bra{\Phi}$ for pure states where the meaning is clear. Equation \eqref{eq:MIPure} is a simple relationship between the range of correlations in $\ket{\Psi}$ and the complexity of preparing it exactly using a unitary circuit.

	The above arguments rely on the premise that the state we wish to prepare is pure. We now discuss the differences that arise when considering mixed states.
	
	\textit{Mixed state complexity.---}We use the following construction to define the complexity of preparing mixed states, which was put forward in Ref.~\cite{Hastings2011}. A mixed state $\rho$ can be generated in depth $d$ if there exists an ensemble of unitaries $\{U_\xx\}$ with corresponding probabilities $p_\xx \geq 0$ such that
	\begin{align}
		\rho &= \sum_\xx p_\xx \ket{\phi_\xx}\bra{\phi_\xx}, \label{eq:DecompSRE} \\ \nonumber 
		\text{ where }  \ket{\phi_\xx} &= U_\xx \ket{0^{\otimes n}}, \; \text{depth}(U_\xx) \leq d \;\; \forall \xx.
	\end{align}
	From any such representation \eqref{eq:DecompSRE}, the algorithm to prepare $\rho$ involves classically sampling from $p_\xx$ and applying the corresponding circuit $U_\xx$. There will generally be many such representations, having different depths $d$, and we define $d_{\rm min}(\rho)$ as the smallest $d$ such that a decomposition \eqref{eq:DecompSRE} exists. Any decomposition of depth $d = d_{\rm min}(\rho)$ will be called `optimal'.  For example, separable (i.e.~unentangled) states $\rho_{\rm SEP}$ by definition can be written as a mixture of product states $\ket{\phi_\xx} = \bigotimes_{j=1}^n \ket{\phi_{\xx,j}}$. Each such $\ket{\phi_\xx}$ can be generated using a product of single-qubit unitaries, which define to have depth zero; thus we have $d_{\rm min}(\rho_{\rm SEP}) = 0$.
	
	Our aim is to derive a lower bound on $d_{\text{min}}(\rho)$ that is analogous to the lower bound on $d_{\text{min}}(\Psi)$ above. However, there are two important technical barriers.  Firstly, mixed states, unlike pure states, can exhibit correlations that are purely classical. For instance, the state $\rho = \frac{1}{2}(\ket{0^{\otimes n}}\bra{0^{\otimes n}} + \ket{1^{\otimes n}}\bra{1^{\otimes n}})$ is separable, and so $d_{\text{min}}(\rho)=1$; 
	nevertheless this state has long-range connected correlations $\braket{Z_iZ_j}_c = 1$ for all pairs of qubits $i,j$. For this reason, the mutual information \eqref{eq:MIPure} (which has been well-characterized in thermal states \cite{Wolf2008, Kuwuhara2021}), cannot be used to lower bound complexity. Secondly, the space of possible pure-state decompositions \eqref{eq:DecompSRE} of a fixed $\rho$ is very large, which makes searching for optimal decompositions an intractable problem at large $n$ \cite{gurvits2004classical,brandao2011quasipolynomial}. Accordingly, our aim will be to construct a lower bound on $d_{\min}(\rho)$ that does not depend on any particular choice of decomposition.
	
	Before continuing, we remark that mixed states can also be generated by adding local ancilla qubits, which are traced out at the end. This allows for a wider range of operations, including non-unital local channels (e.g.~qubit resets). While we do not discuss this possibility in the main text, all our complexity bounds described subsequently still hold if this extra resource is allowed \cite{SM}.

	\textit{Complexity lower bound.---}We now derive our lower bound on $d_{\rm min}(\rho)$, which will generalise the bound described above for pure states [Eq.~\eqref{eq:MIPure}]. As before, the system is divided into three regions $G = A \sqcup B \sqcup C$, with $A$ and $B$ separated by a graph distance $x_{AB}$ (Fig.~\ref{fig:Lightcone}). We will use a proof by contradiction, where we start from the assumption $d_{\rm min}(\rho) < \lowerdist $. This implies that there exists an optimal decomposition \eqref{eq:DecompSRE} where each unitary $U_\xx$ has depth $< \lowerdist$. Using the same lightcone-based arguments that led us to Eq.~\eqref{eq:MIPure}, the mutual information between $A$ and $B$ for each pure state $\ket{\phi_\xx}$ in the optimal decomposition must vanish, $I(A:B)_{\phi_\xx} = 0$.
	
	It is now convenient to define the classical-quantum state associated with this particular decomposition 
	\begin{align}
		\rho^{ABCR}_{\rm opt} \coloneqq \sum_\xx p_\xx   \ket{\phi_\xx}\bra{\phi_\xx}_{ABC} \otimes \ket{\xx}\bra{\xx}_R.
		\label{eq:RhoCQ}
	\end{align}
	Here, $\ket{\xx}_R$ are a set of orthogonal states on a classical register $R$. A key property of such classical-quantum states is that the conditional mutual information $I(A:B|R) \coloneqq I(A:BR) - I(A:R)$ can be expressed as the average over mutual informations in the states $\ket{\phi_\xx}$ \cite{Wilde2013}. Thus we have ${I(A:B|R)_{\rho_{\rm opt}} = \sum_\xx p_\xx I(A:B)_{\phi_\xx} = 0}$. The vanishing of the conditional mutual information (CMI) reflects the fact that all correlations between $A$ and $B$ in the state $\rho^{ABC}$ are due to uncertainty in the classical random variable $\xx$---once we have access to $\xx$, there are no remaining correlations.

	The property just derived cannot yet be applied because it is specific to the optimal decomposition of $\rho^{ABC}$ into a mixture of pure states, which we are unlikely to know in practice.
	To overcome this barrier, we consider the corresponding purification of $\rho^{ABC}$
	\begin{align}
		\ket{\Psi^{ABCE}_\rho} = \sum_\xx \sqrt{p_\xx} \ket{\xx}_E \otimes \ket{\phi_\xx}_{ABC},
		\label{eq:PurificationDef}
	\end{align}
	where $E$ is now a coherent quantum register that is entangled with $ABC$, such that $\Tr_E[\Psi^{ABCE}_\rho] = \rho^{ABC}$. The advantage of working with $\ket{\Psi^{ABCE}_\rho}$ is that all purifications of $\rho^{ABC}$ are related to one another by isometries on $E$, which do not affect von Neumann entropies.  Thus, any entropic inequalities that we derive pertaining to the particular purification $\ket{\Psi^{ABCE}_\rho}$ will also apply if we use a different ensemble of pure states to construct \eqref{eq:PurificationDef}; this circumvents the need to know the optimal decomposition explicitly.
	
	Now, observe that $\rho^{ABCR}_{\rm opt}$ can be generated from $\ket{\Psi^{ABCE}_\rho}$ by dephasing $E$ in the $\ket{\xx}$ basis. This process  can be represented by a quantum channel $\mathcal{E}^{E \rightarrow R}$ on $E$. Using the fact that the mutual information between two subsystems can only decrease when a channel is applied to one of them (the data processing inequality \cite{Wilde2013}), we have
	\begin{align}
		I(A:E)_{\Psi_\rho} \geq I(A:R)_{\rho_{\rm opt}} \!= I(A:BR)_{\rho_{\rm opt}} \geq I(A:B)_{\rho}.\label{eq:inequalitychain}
	\end{align}
	The central equality is due to the vanishing of CMI in the optimal decomposition, while the last inequality follows from the fact that tracing out $R$ is itself a channel on $BR$.
	
	Finally, if an arbitrary channel $\mathcal{N}^{A \rightarrow A'}$ is applied to subsystem $A$, resulting in a modified state $\rho^{A'BCE} = \mathcal{N}^{A \rightarrow A'}[\Psi^{ABCE}_\rho]$, we claim that the analogous chain of inequalities \eqref{eq:inequalitychain} still holds, and in particular $I(A':E) \geq I(A':B)$. This is because the modified classical-quantum state $\rho^{A'BCR}_{\rm opt} = \mathcal{N}^{A \rightarrow A'}[\rho^{ABCR}_{\rm opt}]$ must have vanishing CMI by the data processing inequality, $I(A':B|R) \leq I(A:B|R) = 0$. Since we arrived at this result from a starting assumption that $d_{\rm min}(\rho) < \lowerdist$, the contrapositive gives us our bound.
	\begin{theorem}
		\label{thm:Exact}
		Given a mixed state $\rho^{ABC}$, take any purification $\ket{\Psi^{ABCE}_\rho}$, and apply to it an arbitrary channel $\mathcal{N}^{A \rightarrow A'}$ that acts on $A$. We then have
		\begin{align}
			I(A':B) &> I(A':E) 
			&	\Rightarrow d_{\rm min}(\rho) \geq \lowerdist,
			\label{eq:MIDiffPur}
		\end{align}
		where $x_{AB}$ is the minimum graph distance between $A$ and $B$.
	\end{theorem}
	Equivalently stated, when $I(A':B) > I(A':E)$, the past lightcones of $A$ and $B$ must overlap in some of the unitaries $U_\xx$ used to prepare $\rho^{ABC}$. 
	By applying this result for different choices of subregions $A$,$B$,$C$, and channels $\mathcal{N}^{A \rightarrow A'}$, one can derive lower bounds on the circuit depth needed to exactly prepare $\rho^{ABC}$. Approximate state preparation is dealt with later---see Theorem \ref{thm:Approx}.

	We emphasise again that this result applies irrespective of which decomposition we use to construct the purification.
	To illustrate this, we can re-express the inequality \eqref{eq:MIDiffPur} in terms of entropies of the system only with the help of the \textit{Stinespring dilation} \cite{Wilde2013}---an isometry $W^{A \rightarrow A'A''}$ that reproduces the action of the channel $\mathcal{N}^{A \rightarrow A'}$ when we trace out the complementary system $A''$. Since $\ket{\Psi^{A'A''BCE}_\rho} = W^{A \rightarrow A'A''}\ket{\Psi^{ABCE}_\rho}$ is globally pure, and tracing out $A''$ yields the desired state $\rho^{A'BCE}$, we have
	\begin{align}
		S(B) + S(A''BC) - S(ABC) - S(A'B) > 0 \nonumber\\ \Rightarrow d_{\rm min}(\rho) > \lowerdist.
		\label{eq:SystemEntropies}
	\end{align}
	The left hand side is manifestly a function of $\rho^{ABC}$ and $\mathcal{N}^{A \rightarrow A'}$ only, independent of any choice of decomposition.
	
	\textit{Interpretation.---} By viewing the mixed state $\rho^{ABC}$ as being entangled with `environment' degrees of freedom $E$, our criterion can be thought of as a comparison between correlations within the system $I(A':B)$ and correlations external to the system $I(A':E)$. Our choice to apply a channel $\mathcal{N}^{A \rightarrow A'}$ is  important here: while $I(A':B)$ and $I(A':E)$ are each non-increasing under the application of this channel, the difference $I(A':B)-I(A':E)$ may increase if we choose $\mathcal{N}^{A \rightarrow A'}$ such that it degrades correlations with $E$ more than it does with $B$. For instance, if the region $A$ contains some qubits 
	that are in a maximally mixed state, 
	then when we construct the purification \eqref{eq:PurificationDef}, this will result in a large contribution to $I(A:E)$. By choosing $\mathcal{N}^{A \rightarrow A'}$ such that these qubits are traced out, we can remove this contribution while ensuring that $I(A':B) = I(A:B)$, thus resulting in a net increase of the left hand side of \eqref{eq:MIDiffPur}. 
	
	When $\mathcal{N}^{A \rightarrow A'}$ is identity or isometric, the left hand side of \eqref{eq:SystemEntropies} reduces to the coherent information $I(A\rangle B) \coloneqq S(B) - S(AB)$ \cite{Wilde2013}, which can only be positive if the reduced state $\rho^{AB}$ is entangled. In fact, any measure of bipartite entanglement in $\rho^{AB}$ (e.g.~negativity \cite{vidal2002computable,plenio2005logarithmic}) could be used to construct a similar bound. However, these are unlikely to be useful in typical many-body states, even in the pure-state case. This is because the degrees of freedom in $A$ and $B$ are typically more correlated with their immediate surroundings $C$ than they are with each other, so we expect that $\rho^{AB}$ will generally be separable beyond some constant lengthscale \cite{Parez2023,Parez2024}. In contrast, as we shall see below, if $\mathcal{N}^{A \rightarrow A'}$ is non-isometric, then our diagnostic \eqref{eq:MIDiffPur} depends on the structure of the full state $\rho^{ABC}$, and this can be sensitive enough to obtain nontrivial bounds even if $\rho^{AB}$ is separable.

	\textit{Measurement channels and private correlations.---}Let us consider the special case where $\mathcal{N}^{A \rightarrow A'}$ describes a measurement on $A$, the outcome of which is stored in a classical output register $A'$. Formally, the measurement process is represented by a positive operator-valued measure $\{F_a^A\}$, a set of positive semi-definite operators on $A$ that satisfy $\sum_a F_a^A = I^A$, such that $\mathcal{N}^{A \rightarrow A'}[\,\boldsymbol{\cdot}\,] = \sum_a \Tr[F^A_a \,\boldsymbol{\cdot}\,] \otimes \ket{a}\bra{a}_{A'}$. The probability of obtaining outcome $a$ is $p_a = \Tr[\rho^A F^A_a]$, and the state of the other degrees of freedom conditioned on outcome $a$ is $\rho^{BCE}_a = \Tr_A[F^A_a\Psi^{ABCE}]/p_a$. Equation \eqref{eq:MIDiffPur} then becomes
	\begin{align}
		K \coloneqq \chi_B - \chi_E &> 0 & \Rightarrow  d_{\rm min}(\rho) \geq \lowerdist,
		\label{eq:KeyBound}
	\end{align}
	where $\chi_B = S(\rho^B) - \sum_a p_a S(\rho^B_a)$ is the Holevo information for $B$ (similarly for $\chi_E$) which measures how much the measurement reduces the von Neumann entropy of $B$ on average. As discussed in Ref.~\cite{Garratt2025}, the quantity $K$ is a special instance of the \textit{private information} \cite{Wilde2013}, that arises in the private classical capacity theorem of Devetak, Cai, Winter, and Yeung \cite{Devetak2005a,Devetak2005,Cai2004}. This quantity has an operational interpretation as the rate at which quantum key distillation between $A$ and $B$ can be achieved, using the state $\rho^{ABC}$ as a resource.  
	
	To summarise Eq.~\eqref{eq:KeyBound}: if a measurement on $A$ purifies $B$ more than it does $E$ (as measured by the Holevo information), then $\rho^{ABC}$ contains correlations between $A$ and $B$ that are private from $E$. This implies an overlap of past lightcones of $A$ and $B$, and hence a circuit depth lower bound $d_{\rm min}(\rho^{ABC}) > \lowerdist$.
	
	\textit{Weak measurements and thermal states.---}While versatile, the information-theoretic quantities we have been working with so far may be challenging to compute or interpret in many-body systems. To simplify the expressions, it is helpful to consider the limit of weak measurements, following the approach that we developed in Ref.~\cite{Garratt2025}. We restrict to two-outcome POVMs $a = 0, 1$ of the form $F_a^A = q_aI^A + (-1)^a\mu O^A$, where $q_{0,1}$ are positive numbers satisfying $q_0 + q_1 = 1$, $O^A$ is a bounded operator, and $\mu$ quantifies the measurement strength. Both $\chi_B$ and $\chi_E$ vanish up to first order in $\mu$, so we can look at the second order contributions $K^{(2)} = \chi_B^{(2)} - \chi_E^{(2)}$, where
	\begin{align}
		\chi_{E}^{(2)} &\coloneqq \frac{1}{2}\left.\frac{ \partial^2 \chi_{E}}{\partial \mu^2}\right|_{\mu = 0}.
		\label{eq:HolevoDeriv}
	\end{align}
	If $K^{(2)} > 0$, then we infer the same circuit depth lower bound $d_{\rm min}(\rho) \geq \lowerdist$. For completeness, general expressions for $\chi_{B,E}^{(2)}$ are given in the appendix. To provide intuition, here we specialise to Gibbs states, $\rho_\beta = e^{-\beta H}/\Tr[e^{-\beta H}]$ of some local Hamiltonian $H$. As shown in Ref.~\cite{Garratt2025}, $\chi_E^{(2)}$ can be related to dynamical properties of the Hamiltonian $H$ at the specified temperature $\beta^{-1}$. Specifically, we have $\chi^{(2)}_E = \frac{1}{4\pi}\int_{-\infty}^{\infty} \dif \omega C_\beta(O^A, \omega) f_\beta(\omega)$, where $C_\beta(O^A, \omega)$ is the Fourier transform of the connected dynamical correlation function $C_\beta(O^A, t) = \Tr[O^A(t) O^A(0) \rho_\beta] - \Tr[O^A \rho_\beta]^2$, and $f_\beta(\omega) = \frac{\beta \omega}{(e^{\beta \omega}-1)}$. Moreover, one can straightforwardly lower bound $\chi_B^{(2)}$ in terms of connected spatial correlation functions, $\chi_B^{(2)} \geq \braket{O^AO^B}_c^2$ for any observable $O_B$ that has unit norm and is supported in $B$. Thus, just from knowledge of the spatial and temporal correlations in thermal equilibrium, we can infer lower bounds on the circuit depth required to prepare the thermal state. This particular instantiation of our bound will prove very useful when studying the complexity of Gibbs states of critical Hamiltonians later on.
	
	\textit{Approximate state preparation.---}So far, we have focused on the complexity of exact state preparation. However, it is conceptually and practically important to consider the resources required to preparing an approximation of $\rho$ of some specified error tolerance $\epsilon$. For instance, trivial ground states with exponentially decaying correlations would technically require a deep circuit to reproduce the extremely small correlations between distant qubits; yet, a constant depth suffices to prepare the state approximately \cite{Hastings2005, Chen2010}. Accordingly, we write $d_{\rm min}(\rho, \epsilon)$ for the minimum depth required to prepare a state $\rho'$ that is $\epsilon$-close to the target state, as measured by the trace distance $T(\rho, \rho') = \frac{1}{2}\|\rho - \rho'\|_1$, with $\|X\|_1 = \Tr[\sqrt{X^\dagger X}]$ the trace norm \footnote{In the terminology of Ref.~\cite{Hastings2011}, we say that $\rho$ is $(d, \epsilon)$-trivial for any $d \geq d_{\rm min}(\rho, \epsilon)$.}. For $\epsilon = 0$, $d_{\rm min}(\rho, \epsilon=0)$ is the depth of the optimal decomposition that appeared in Eq.~\eqref{eq:RhoCQ}. By establishing continuity bounds for the relevant quantities that appear in the inequalities proved above, we can obtain a lower bound on $d_{\rm min}(\rho, \epsilon)$.
	\begin{theorem}
		Given a mixed state $\rho^{ABC}$, take any purification $\ket{\Psi^{ABCE}}$, and apply to it an arbitrary channel $\mathcal{N}^{A \rightarrow A'}$ that acts on $A$. There is a function $k(\epsilon)$ such that $I(A':B)_\rho - I(A':E)_\rho > k(\epsilon)$ implies that $d_{\rm min}(\rho, \epsilon) \geq \lowerdist$. When $\mathcal{N}^{A \rightarrow A'}$ is a weak measurement, the same conclusion can be made if $\chi_B^{(2)} - \chi_E^{(2)} > 12 \epsilon$. These bounds still apply if local ancilla qubits are allowed when preparing $\rho^{ABC}$.
		\label{thm:Approx}
	\end{theorem}
	Theorem \ref{thm:Approx} is proved in the supplement \cite{SM}, where we also specify the function $k(\epsilon)$, which behaves as $O(\epsilon \log(1/\epsilon))$ for small $\epsilon$.

	\begin{figure}
		\includegraphics[width=0.47\textwidth]{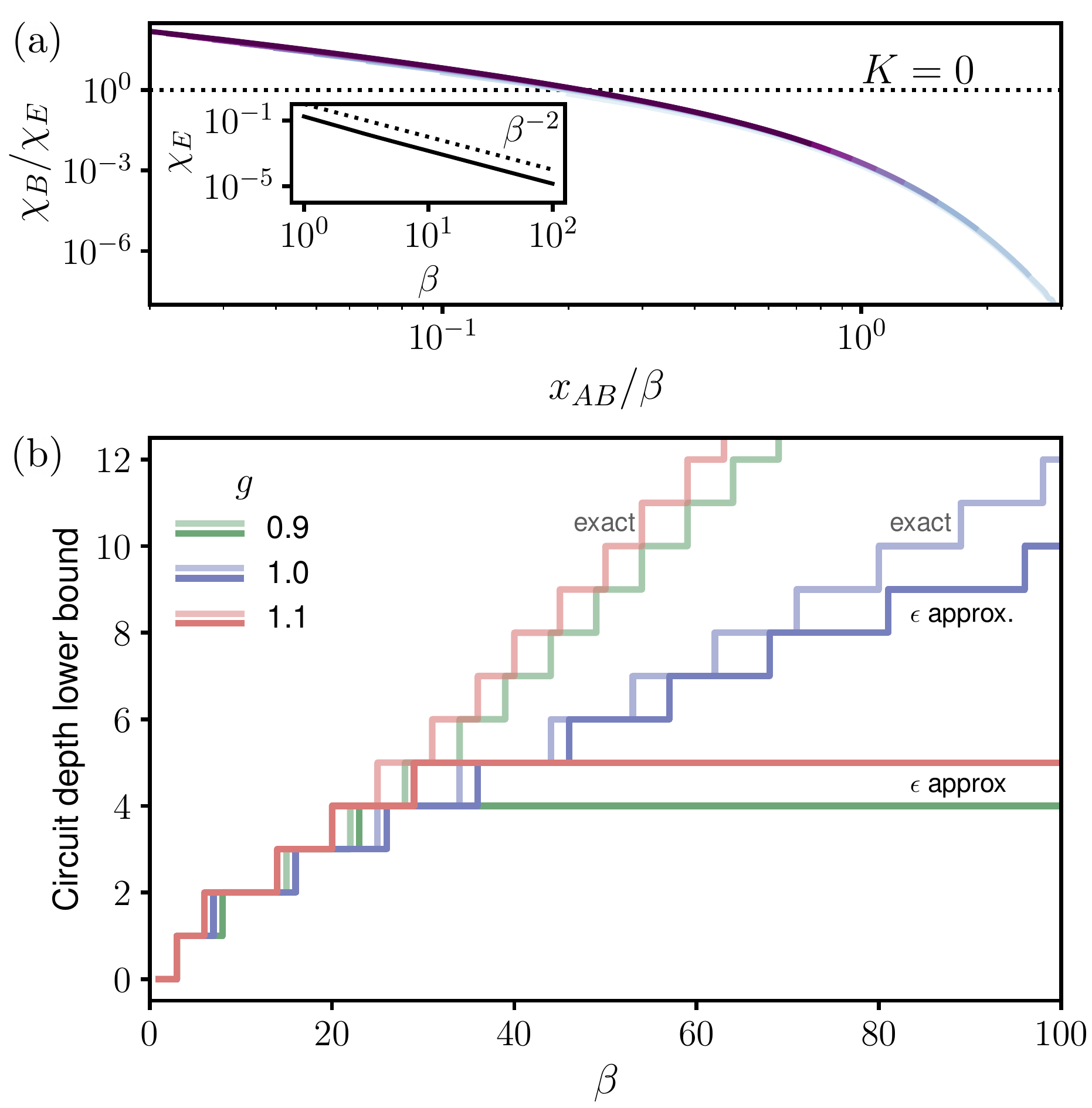}
		\caption{
			Lower bounds on the circuit depth needed to prepare thermal states $\rho \propto e^{-\beta H}$ of the transverse field Ising chain, $H = -\sum_j( Z_j Z_{j+1} + g X_j)$. Here we use open boundary conditions, the length $n$ of the chain is $n=301$, $A$ is the central site, and $B$ consists of the first $(n+1)/2-x_{AB}$ sites of the chain. The channel $\mathcal{N}^{A \to A'}$ is a projective measurement of the Pauli-$X$ operator, so Eq.~\eqref{eq:KeyBound} applies. (a) Ratio $\chi_B/\chi_E$ at the quantum critical point, $g=1$, for various $x_{AB}$ and $\beta=10,20,\ldots,100$ (light to dark). 
			Inset: Decay of $\chi_E$ (solid) at $g=1$ on increasing $\beta$, compared with the expected behavior $\sim \beta^{-2}$ (dotted). (b) Lower bounds on circuit depth for (light) exact and (dark) $\epsilon$-approximate state preparation for various transverse field strength $g$ (different colors, see legend). The bounds for approximate state preparation come from the condition $K \leq k(\epsilon)$ with $k(\epsilon) = 10^{-5}$, corresponding to $\epsilon \approx 10^{-7}$.
		}
		\label{fig:Ising}
	\end{figure}
	
	\textit{Thermal states of critical Hamiltonians.---}Having derived our various circuit depth lower bounds, we now  apply them to understand the complexity of various many-body systems of interest. Here we will focus on thermal states of one-dimensional Hamiltonians near a quantum critical point.
	
	We start by taking a critical Hamiltonian $H$ on an infinite chain, whose low-energy and long-wavelength behaviour is described by a $(1+1)$-dimensional CFT. 
	Employing the formalism described above, we consider a weak measurement of an observable $O^A$ that couples to some primary field at the origin $x = 0$ (see e.g.~Refs.~\cite{Garratt2023,Murciano2023} for previous treatments of weak measurements in CFTs). The region $B$ is chosen to be the semi-infinite line $x \in [x_{AB},\infty]$, such that regions $A$ and $B$ are separated by a distance $x_{AB} > 0$ on a lattice with unit spacing. Both $\chi^{(2)}_E$ and $\chi^{(2)}_B$ can be computed within the CFT. In the appendix, we show that this results in a lower bound on the circuit depth required to prepare a thermal state to accuracy $\epsilon$ of
	\begin{align}
		d_{\rm min}(\rho^{\rm critical}_\beta, \epsilon) \geq \frac{\beta}{4\pi}\ln\left(1 + \frac{1}{(1 + c\beta^\eta \epsilon)^{1/\eta}} \right),
		\label{eq:CFTLowerBound}
	\end{align}
	where $c>0$ is a non-universal constant, and $\eta = 2\Delta_{\rm min}$ depends on the smallest scaling dimension $\Delta_{\rm min}$ of all primary fields in the CFT. This expression is reliable only in the low-temperature limit $\beta \gg 1$, where CFT provides an approximate description of the correlations in a critical lattice model.
	
	The behaviour in \eqref{eq:CFTLowerBound}---in particular the divergence as $\beta$ is increased---can be understood in a way that generalizes beyond weak measurements. The only length- and time-scales in the problem are $x_{AB}$ and $\beta$ so, for any local channel $\mathcal{N}^{A \to A'}$, by conformal invariance we have $\chi_B/\chi_E = \gamma(x_{AB}/\beta)$ for $x_{AB},\beta \gg 1$ and a channel-dependent dimensionless function $\gamma(u)$. By the data processing inequality, $\gamma(u)$ is non-increasing. 
	For exact state preparation, Eq.~\eqref{eq:KeyBound} gives $d_{\rm min}(\rho_\beta) \geq \beta u^\star$, where $u^\star$ is the point where $\gamma(u^\star) = 1$ \footnote{Since $\chi_E \to 0$ for $\beta \to 0$, $\gamma(u\to \infty)=0$, and since $\chi_B \to 0$ for $x_{AB}\to \infty$, $1/\gamma(u \to 0)=0$. Therefore there must exist a $u^*$ such that $\gamma(u^\star) = 1$}. From Theorem \ref{thm:Approx}, for approximate state preparation $u^\star$ should be replaced with $u^\star_\epsilon$, the solution of $\gamma(u^\star_\epsilon) = 1 + k(\epsilon)/\chi_E$. Assuming that $\gamma(u)$ is analytic and non-stationary in the vicinity of $u^\star \equiv u^\star_0$, 
	a Taylor expansion then gives the lower bound on circuit depth $d_{\rm min}(\rho_\beta^{\rm critical},\epsilon) \geq \beta\frac{u^\star_0}{2}(1-\kappa_\beta k(\epsilon) )$ in the small-$\epsilon$ limit, where $\kappa^{-1}_\beta  = -\chi_E(\beta)\times\partial_u\gamma(u)|_{u=u^\star} \geq 0$. This is consistent with \eqref{eq:CFTLowerBound}, using the fact that $\chi_E \propto \beta^{-\eta}$ for weak measurements \cite{SM}.  
	
	In Fig.~\ref{fig:Ising}(a) we verify this prediction numerically for thermal states of the transverse field Ising chain $H = -\sum_j (Z_j Z_{j+1}+g X_j)$ at its quantum critical point, $g=1$, and for $\mathcal{N}^{A \to A'}$ a projective measurement of a local $X$ operator. The ratio $\chi_B/\chi_E$ indeed collapses onto a function of $x_{AB}/\beta$, and our results indicate $\frac{u_0^*}{2} \approx 0.1$ for this measurement [larger than the coefficient $\frac{\ln 2}{4\pi}$ for weak measurements \eqref{eq:CFTLowerBound}]. The $X$ operator 
	couples to a primary field with $\Delta_{\text{min}}=1$, so $\chi_E \sim \beta^{-2}$ as we confirm in the inset. In Fig.~\ref{fig:Ising}(b), 
	we plot the resulting lower bounds on circuit depth,  both at and away from the critical point. 
	For exact preparation ($\epsilon=0$), $d_{\rm min}(\rho_\beta)$ diverges with increasing $\beta$ for all choices of $g$. For approximate state preparation ($\epsilon > 0$), $d_{\rm min}(\rho_\beta,\epsilon)$ diverges only at the critical point $g = 1$. Away from criticality $g \neq 1$, there is a plateau beyond some moderate $\beta$, which reflects the finite correlation length at $\beta \rightarrow \infty$. This highlights the importance of considering approximate state preparation: The lower bound for exact preparation is highly sensitive to the exponentially decaying, but nonzero, correlations at $g \neq 1$.

	\textit{Discussion.---}We have established a new method for characterising mixed state complexity in terms of correlations (Theorems \ref{thm:Exact} and \ref{thm:Approx}), and applied our results to Gibbs states of quantum critical Hamiltonians. This represents an important early step in understanding the landscape of complexity in mixed quantum states, which has remained much less explored than the pure-state case. Our analysis complements recent results showing that Gibbs states of local Hamiltonians with sufficiently high temperature $\beta < \beta^\star$ can be prepared with single-qubit gates, $d_{\rm min}(\rho_{\beta < \beta^{\star}}) = 0$ \cite{Bakshi2025}.
	
	The depth of the ensemble of unitaries $U_\xx$ needed to prepare a mixed state $\rho$ [via Eq.~\eqref{eq:DecompSRE}] is a natural measure of time complexity for quantum computers that operate with unitary gates of fixed connectivity. However, in other settings, one may wish to remove the geometric locality condition on the preparation circuits. Since our method allows one to identify  pairs of qubits that must have overlapping past lightcones, it could in principle be adapted to lower bound the required gate count in all-to-all circuits. Another way in which the allowed computational resources can be enhanced is to allow mid-circuit measurements and feed-forward, which can significantly speed up state preparation  \cite{Briegel2001,Piroli2021,Bravyi2022,Malz2024,Tantivasadakarn2024}. It would be interesting to extend our arguments to include these operations, and more generally to understand the limits imposed by causality on the propagation of quantum and classical correlations in adaptive circuits \cite{friedman2023locality, lu2023mixed, liu2025state}.
	
	Finally, we anticipate that our results will be useful in the ongoing effort to characterize topological order in mixed states \cite{Hastings2011,Fan2024,Chen2024,Wang2025,Zhou2025}. Rigorous lower bounds on circuit depth that depend only on entropic quantities have proved elusive even for pure states, since common diagnostics such as topological entanglement entropy \cite{Kitaev2006, Levin2006} can be spoofed by trivial states \cite{Zou2016}. Nevertheless, we expect that our method of invoking purifications could be used to adapt other robust measures of topological order from the pure to mixed state setting.
	
	\textit{Acknowledgements.---} The authors are grateful to Matteo Ippoliti, Michael Levin, and Thomas Schuster for useful discussions. M.M.~acknowledges support from Trinity College, Cambridge. S.J.G.~was supported by the Gordon \& Betty Moore Foundation, and in part by a Brown Investigator Award, a program of the Brown Institute for Basic Sciences at the California Institute of Technology.

	\bibliography{circuit_depth.bib}

	\begin{onecolumngrid}
		\newpage
		
		\begin{center}
			{\fontsize{11}{11}\selectfont
				\textbf{Supplemental Material for  ``\papertitle''\\[5mm]}}
			{\normalsize Max McGinley and Samuel J.~Garratt \\[1mm]}
			
		\end{center}
		\normalsize\

		\setcounter{equation}{0}
		\setcounter{figure}{0}
		\setcounter{table}{0}

		\renewcommand{\theequation}{S\arabic{equation}}
		\renewcommand{\thefigure}{S\arabic{figure}}
		\renewcommand{\thesection}{S\arabic{section}}
		
		\section{Holevo information for weak measurements}
		
		In this appendix we provide a general expression for the leading order contributions to the Holevo information, defined in the main text as Eq.~\eqref{eq:HolevoDeriv}. A proof of these formulae can be found in the appendix of Ref.~\cite{Garratt2025}.
		
		As specified in the main text, we consider a two-outcome POVM $F_a^A = \frac{1}{2}(I^A + (-1)^a\mu O_a)$, where $a = 0,1$ labels the possible outcomes, $O^A$ is a Hermitian operator of bounded norm $\|O^A\|_\infty = 1$, and $\mu$ is the measurement strength. We use $X$ as a placeholder for either region $B$ or $E$, as appropriate. Then, defining the corresponding operator $\xi^X \coloneqq  \Tr_{ABCE \backslash X}[O^A \ket{\Psi^{ABCE}}\bra{\Psi^{ABCE}}]$, we have
		\begin{align}
			\chi^{(2)}_X = \frac{1}{2}\left(\Tr\Big[\xi^X \mathcal{T}_{\rho^X}[\xi^X]\Big] - \Tr[\rho^A O^A]^2\right)
			\label{eq:Chi2General}
		\end{align}
		where, following Ref.~\cite{Lieb1973}, for any given density matrix $\sigma$ we define the corresponding completely positive map
		\begin{align}
			\mathcal{T}_\sigma[\xi] &\coloneqq \int_0^\infty \dif z \frac{1}{\sigma + zI}\xi\frac{1}{\sigma + zI}  \label{eq:TDef} \\
			&= \sum_{ij} \braket{i|\xi|j} \frac{\ln \lambda_i - \ln \lambda_j}{\lambda_i - \lambda_j} \ket{i}\bra{j}.
		\end{align}
		In the second line, we use $\{(\lambda_i, \ket{i})\}_i$ to denote eigenvalue-eigenvector pairs of the state $\sigma$. Note that for $X = E$, the state $\rho^E$ has the same eigenvalues as the state of the system $\rho^{ABC}$. Thus, for Gibbs states $\rho^{ABC} \propto e^{-\beta H}$, we can work in the eigenbasis of the Hamiltonian $H\ket{E} = E\ket{E}$, giving (see Ref.~\cite{Garratt2025})
		\begin{align}
			\chi^{(2)}_E &= \sum_{E,E'} e^{-\beta(E+E')} |\braket{E|O^A|E'}|^2 \frac{\beta(E-E')}{e^{\beta E}-e^{\beta E'}} \nonumber\\
			&= \frac{1}{4\pi} \int_{-\infty}^\infty \dif \omega C_\beta(O_A,\omega) f_\beta(\omega)
			\label{eq:ChiEThermal}
		\end{align}
		where $C_\beta(O_A,\omega)$ is the frequency-domain autocorrelation function of $O^A$, i.e.~the Fourier transform of $C_\beta(O_A,t) = \braket{O^A(t)O^A(0)}_{\beta, c}$.
		
		Since $\chi_B$ can be viewed as the mutual information between the classical random variable $a$ and the quantum state of $B$, the data processing inequality applies if we apply a channel to $B$. This gives us a way to lower bound $\chi^{(2)}_B$. In particular, we can apply another measurement to $B$, represented by the two-outcome POVM $F^B_b = \frac{1}{2}(I^B + (-1)^b O^B)$ for some observable $O^B$ of bounded norm $\|O^B\|_\infty \leq 1$ (note that this is not a weak measurement). The mutual information between the measurement outcomes $a$ and $b$ then serves as a lower bound for $\chi_B$, and in particular the leading order contributions can be calculated using the expression \eqref{eq:Chi2General}, giving
		\begin{align}
			\chi^{(2)}_B \geq \frac{\braket{O^AO^B}_{c, \rho^{AB}}^2}{1 - \braket{O_B}^2_{\rho^B}}
		\end{align}
		where the connected correlator is $\braket{O^AO^B}_{c, \rho^{AB}} = \Tr[O^AO^B \rho^{AB}] - \Tr[O^A\rho^A]\Tr[O^B\rho^B]$, as in the main text.
		
		\section{Allowing for local ancilla qubits}

		Here we justify the claim made in the main text that our lower bounds on the circuit depth required to prepare $\rho^{ABC}$ still hold if local ancilla qubits are allowed in the preparation circuit. While the structure of local circuits with ancillas has been studied before, to keep our discussion self-contained is helpful to sharpen what extra resources we are allowing for here with the following definition.
		
		\newtheorem{definition}{Definition}
		\begin{definition}[Local ancilla-assisted circuits]
			For an $n$-qubit system on a graph $G$ with Hilbert space $\mathcal{H} = \bigotimes_{v \in G} \mathcal{H}_v$, a state $\rho$ can be prepared by an ensemble of local ancilla-assisted circuits of depth $d$ if the following construction exists. A finite-dimensional ancilla qudit with Hilbert space $\mathcal{A}_v$ is associated to each vertex $v$, and an ensemble of circuits $\{(p_\xx, W_\xx)\}_\xx$ is chosen such that
			\begin{align}
				\rho = \sum_\xx p_\xx \Tr_\mathcal{A}\left[ W_\xx \left(\bigotimes_{v \in G} \left(\ket{0}\bra{0}_{\mathcal{H}_v} \otimes \ket{0}\bra{0}_{\mathcal{A}_v}\right)\right) W_\xx^\dagger \right].
				\label{eq:AncillaPrep}
			\end{align}
			where $\ket{0}_{\mathcal{A}_v}$ is a reference in initial state for the ancilla qudit at $v$, and $\mathcal{A} = \bigotimes_{v \in G} \mathcal{A}_v$. Each $W_\xx$ is a circuit composed of unitary gates that act on the system qubits and ancilla qudits on two vertices that share an edge in $G$, and the depth of each $W_\xx$ is no greater than $d$.
			\label{def:Ancilla}
		\end{definition}
		This allows for a broader range of operations than ensembles of local unitaries alone. For instance, as a consequence of Stinespring's dilation theorem \cite{Wilde2013}, ancillae can be used to simulate arbitrary local quantum channels, including those that are non-unital (e.g.~the reset channel). Thus, this definition incorporates the set of states that can be prepared by ensembles of non-unitary circuits $\{(p_\xx, \mathcal{E}^{ABC}_\xx)\}_\xx$ where each $\mathcal{E}^{ABC}_\xx$ is composed of local two-qubit quantum channels, with the total depth being $d$.
		
		We wish to highlight that this constructions equivalent to Definition \ref{def:Ancilla} have appeared before in both few-body \cite{Forster2009,Gallego2012,Buscemi2012, Schmid2023understanding} and many-body \cite{Hastings2011, lessa2025higher} settings. When the unitaries $W_\xx$ are restricted to be strictly local, the above can be viewed as the set of free operations in the resource theory of \textit{nonlocality} \cite{Forster2009,Gallego2012,Buscemi2012,Schmid2023understanding}, which encompasses local operations and shared randomness (LOSR) \cite{Buscemi2012}. This contrasts with the resource theory of entanglement, where the free operations are local operations and classical communication (LOCC). In the many-body setting, if one classifies phases through finite-depth LOCC circuits, then topologically ordered states such as the toric code become trivial \cite{Piroli2021}; in contrast, classifying phases according to finite-depth LOSR circuits appears to be more consistent with what is known for the usual definition of phases of matter for pure states \cite{Hastings2011, lessa2025higher}.  
		
		\subsection{Generalising Theorem \ref{thm:Exact} to ancilla-assisted circuits}
		
		Starting from Theorem \ref{thm:Exact}, we will now derive a corresponding lower bound on the minimum depth required to prepare a mixed state $\rho$ with the assistance of local ancillae. As usual, we divide system qubits into $A$, $B$, $C$, and write $\rho \equiv \rho^{ABC}$. When describing states in the extended Hilbert space that includes system qubits and ancilla qudits, we will use $S$ for the former and $T$ for the latter. Subsets of system and ancilla degrees of freedom that are associated with the regions $A$, $B$, $C$ will be denoted $S_{A,B,C}$ and $T_{A,B,C}$, respectively.
		
		For any ensemble of circuits $\{(p_\xx, W_\xx)\}$ with ancilla qudit Hilbert spaces $\{\mathcal{A}_v\}_{v \in G}$ that prepares $\rho$, we define the pure state
		\begin{align}
			\ket{\Phi^{STE}} = \sum_\xx \sqrt{p_\xx} \ket{\xx}_E \otimes W_\xx\ket{0}_{ST}
			\label{eq:PurificationAncilla}
		\end{align}
		By tracing out $E$ we obtain an enlarged mixed state $\sigma^{ST} \coloneqq \Tr_E \Phi^{STE}$ which is supported on both the system and ancillas. By Eq.~\eqref{eq:AncillaPrep} the reduced state on the system $\Tr_T \sigma^{ST} = \Tr_{TE} \Phi^{STE} = \rho^S$, where $\rho^S \equiv \rho^{ABC}$ is the target state on the system qubits.
		
		If we now view the system and ancilla degrees of freedom together as a single enlarged system, we can interpret $\{(p_\xx, W_\xx)\}_\xx$ as an (unassisted) ensemble of unitaries that prepares the enlarged mixed state $\sigma^{ST}$. We can then invoke Theorem \ref{thm:Exact} to put constraints on the depth of the circuits $W_\xx$. Specifically, if we apply any channel $\mathcal{M}^{S_AT_A \rightarrow A'}$ to the state \eqref{eq:PurificationAncilla}, then for the resulting state $\sigma^{A'S_BT_BS_CT_CE} \coloneqq \mathcal{M}^{S_AT_A \rightarrow A'}[\Phi^{STE}]$ we have
		\begin{align}
			I(A':S_BT_B)_\sigma &> I(A':E)_\sigma \quad \Rightarrow\quad \max_\xx[\text{depth}(W_\xx)] \geq \lowerdist
			\label{eq:EnlargedLowerBound}
		\end{align}
		We now let $\mathcal{N}^{S_A \rightarrow A'}$ be an arbitrary channel that takes only the system qubits in $A$ as input, and define $\mathcal{M}^{S_AT_A \rightarrow A'} = \mathcal{N}^{S_A \rightarrow A'} \otimes \mathcal{T}^{T_A \rightarrow \emptyset}$, where $\mathcal{T}^{T_A \rightarrow \emptyset}[\,\cdot\,] = \Tr_{T_A}[\,\cdot\,]$ is a channel that traces out the input and returns a trivial (scalar) output. The state $\sigma^{A'S_BT_BS_CT_CE}$ that results from applying this channel to $\ket{\Phi^{STE}}$ has a reduced state on $A'S_BS_C$ that is equal to the perturbed target state $\rho^{A'BC} \coloneqq \mathcal{N}^{A \rightarrow A'}[\rho^{ABC}]$ (with subregions $A$,$B$,$C$ in $\rho$ identified with the corresponding sets $S_{A,B,C}$). By the data processing inequality, we have
		\begin{align}
			I(A':S_BT_B)_{\sigma} \geq I(A':S_B)_{\sigma} = I(A':B)_\rho.
		\end{align}
		Finally, we observe that $\ket{\Phi^{STE}}$ can be viewed as a purification of $\rho^S \equiv \rho^{ABC}$, with $TE$ the purifying system. Then, using the fact that different purifications are related by isometries, along with the data processing again we find
		\begin{align}
			I(A':E)_\rho = I(A':TE)_\sigma \geq I(A':E)_{\sigma}
		\end{align}
		We thus see that  $I(A':B)_\rho$ is a lower bound for the left hand side of the first inequality in \eqref{eq:EnlargedLowerBound}, while $I(A':E)_\rho$ is an upper bound for the right hand side. So, bringing everything together and recalling Eq.~\eqref{eq:EnlargedLowerBound} we have
		\begin{align}
			I(A':B)_\rho > I(A':E)_\rho \quad \Rightarrow \quad I(A':S_BT_B)_{\sigma} \geq I(A':E)_{\sigma} \quad \Rightarrow \quad  \max_\xx[\text{depth}(W_\xx)] \geq \lowerdist
		\end{align}
		This confirms that the circuit depth lower bound proved in Theorem \ref{thm:Exact} continues to hold if we allow for local ancillas, as per Definition \ref{def:Ancilla}.

		\section{Proof of Theorem \ref{thm:Approx} \label{app:Gaussian}}
		
		We quote the full version of Theorem \ref{thm:Approx} here\\
		
		\noindent \textbf{Theorem \ref{thm:Approx} (Formal version)} \emph{Let $\mathcal{N}^{A \rightarrow A'}$ be a channel whose output Hilbert space dimension is $d_{A'}$, and define $k(\epsilon) = 2\epsilon \ln d_{A'} + 4g(\epsilon)$, where $g(x) = (1+x)\ln(1+x) - x\ln x$. Then, given a mixed state $\rho^{ABC}$, take an arbitrary purification $\ket{\Psi^{ABCE}_\rho}$ and apply $\mathcal{N}^{A \rightarrow A'}$ on $A$. If the resulting state satisfies $I(A':B) - I(A':E) > k(\epsilon)$, then the shortest geometrically local circuit that prepares a state that is $\epsilon$-close to $\rho^{ABC}$ in trace distance has depth $d_{\rm min}(\rho^{ABC},\epsilon) \geq \lowerdist$. Similarly, given a weak measurement on $A$  of strength $\mu$, the same lower bound on $d_{\rm min}(\rho^{ABC},\epsilon) \geq \lowerdist$ applies to any state for which $\chi_B^{(2)} - \chi_E^{(2)} > 12 \epsilon$, where $\chi_{B,E}^{(2)}$ are the contributions to the Holevo information at second order in $\mu$ [Eq.~\eqref{eq:HolevoDeriv}]. The same lower bounds on the required circuit depth needed to approximately prepare $\rho^{ABC}$ also hold if local ancilla qubits are used in the circuit as per Definition \ref{def:Ancilla}.}\\
		
		To prove the above statement, we will make use of the circuit depth lower bounds for exact state preparation that we derived in the main text, Eqs.~(\ref{eq:MIDiffPur}, \ref{eq:KeyBound}). To convert these into statements about approximate state preparation, we will need to use the following continuity properties for the quantities that appear in those bounds.
		\begin{lemma}
			\label{thm:Continuity}
			Let $\rho^{ABC}$ and $\sigma^{ABC}$ be two states with purifications $\Psi^{ABCE}_\rho$ and $\Psi^{ABCE}_\sigma$, respectively, whose trace distance satisfies $\frac{1}{2}\|\rho^{ABC} - \sigma^{ABC}\|_1 \leq \epsilon$. For a channel $\mathcal{N}^{A \rightarrow A'}$ with output dimension $d_{A'}$, we define the state $\rho^{A'BCE} = (\mathcal{N}^{A \rightarrow A'}\otimes \textup{id}^{BCE})[\ket{\Psi_\rho^{ABCE}}\bra{\Psi_\rho^{ABCE}}]$, and similarly for $\sigma^{A'BCE}$. Then we have
			\begin{align}
				\big|(I(A':B)_{\rho} - I(A':E)_{\rho})- (I(A':B)_{\sigma} - I(A':E)_{\sigma})\big| \leq k(\epsilon) \coloneqq 2\epsilon \ln d_{A'}   + 4g(\epsilon),
				\label{eq:KeyContFull}
			\end{align}
			where $g(x) \coloneqq (1+x)\ln(1+x) - x \ln x$. Furthermore, for a weak measurement on $A$ of strength $\mu$, the second order contributions to the Holevo quantity [Eq.~\eqref{eq:HolevoDeriv}] satisfy
			\begin{align}
				\left|\chi^{(2)}_E[\rho] - \chi^{(2)}_E[\sigma]\right| \leq 6\epsilon,
				\label{eq:KeyContDeriv}
			\end{align}
			and similarly for $\chi^{(2)}_B$.
		\end{lemma}
		From the above inequalities, we can prove Theorem \ref{thm:Approx} straightforwardly by contradiction. 
		Let us suppose that there is a state $\sigma^{ABC}$ that is $\epsilon$-close to the target state $\rho^{ABC}$, and can be generated  by unitaries of depth at most $x_{AB}/2$. By  Eq.~\eqref{eq:MIDiffPur}, we must have $I(A':B)_\sigma - I(A':E)_\sigma \leq 0$. However, if we combine Eq.~\eqref{eq:KeyContFull} with the assumption in Theorem \ref{thm:Approx} that $I(A':B)_\rho - I(A':E)_\rho > k(\epsilon)$, then we find
		\begin{align}
			I(A':B)_\sigma - I(A':E)_\sigma > I(A':B)_\rho - I(A':E)_\rho - k(\epsilon) \geq 0.
		\end{align}
		which is a contradiction. Thus, no such $\sigma^{ABC}$ can exist. The same logic using Eqs.~(\ref{eq:KeyBound}, \ref{eq:KeyContDeriv}) instead of Eqs.~(\ref{eq:MIDiffPur}, \ref{eq:KeyContFull}) proves the second part of Theorem \ref{thm:Approx}. Finally, as proved in the previous section, the circuit depth lower bound from Theorem \ref{thm:Exact} still holds if we allow for local ancilla-assisted circuits, and this then confirms the final statement. \hfill $\square$

		\subsection*{Proof of Lemma \ref{thm:Continuity}}

		We now prove the two parts of Lemma \ref{thm:Continuity} in sequence. The first part is simpler, and mainly makes use of a continuity bound for the conditional entropy $S(X|Y) \coloneqq S(XY) - S(Y)$ two arbitrary subsystems $X, Y$ of Hilbert space dimension $d_{X,Y}$, respectively. Lemma 2 of Ref.~\cite{Winter2016} states that $|S(X|Y)_\rho - S(X|Y)_\sigma| \leq 2 t \ln d_{X} + g(t)$, where $t = \frac{1}{2}\|\rho^{XY} - \sigma^{XY}\|_1$ and $g(x)$ is the function quoted in Lemma \ref{thm:Continuity}. Using the Stinespring isometry $W^{A \rightarrow A'A''}$ of the channel $\mathcal{N}^{A \rightarrow A'}$ to define the state $\ket{\Psi_\rho^{A'A''BCE}} = (W^{A \rightarrow A'A''}\otimes I^{BCE})\ket{\Psi^{ABCE}_\rho}$, we can rewrite the quantity in question as
		\begin{align}I(A':B)_\rho - I(A':E)_\rho &= S(B)_\rho + S(A''BC)_\rho - S(A'A''BC)_\rho - S(A'B)_\rho \nonumber\\
			&= - S(A'|A''BC) - S(A'|B)
			\label{eq:CondEntDiff}
		\end{align}
		The right hand side of \eqref{eq:CondEntDiff} only depends on the marginal state $\rho^{A'A''BC}$, which is related to $\rho^{ABC}$ by the isometry $W^{A \rightarrow A'A''}$. Since the trace norm is invariant under isometries, and by assumption $\frac{1}{2}\|\rho^{ABC} - \sigma^{ABC}\|_1 \leq \epsilon$, we therefore have $\frac{1}{2}\|\rho^{A'A''BC} - \sigma^{A'A''BC}\|_1 \leq \epsilon$. Applying Lemma 2 of Ref.~\cite{Winter2016} twice and using the triangle inequality, we arrive at Eq.~\eqref{eq:KeyContFull}.\\
		
		To prove the second part of the theorem, we need a corresponding continuity bound for the second derivative of the Holevo information with respect to the measurement strength $\mu$. In the following, $X$ should be viewed as a placeholder for either $E$ or $B$.
		\begin{lemma}
			\label{lem:HolevoContDeriv}
			For two bipartite states $\rho^{AX}$ and $\sigma^{AX}$ that satisfy $\frac{1}{2}\|\rho^{AX} - \sigma^{AX}\|_1 \leq \epsilon$, and a POVM $F_a^A(\mu) = q_a(I + \mu O_a)$  of measurement strength $\mu$ that acts on $A$, the leading order in $\mu$ contributions to the Holevo information $\chi^X$ differ by at most
			\begin{align}
				\left|\chi^{(2)}_X(\rho^{AX}) - \chi^{(2)}_X(\sigma^{AX})\right| \leq 6\epsilon 
				\label{eq:HolevoContDeriv}
			\end{align}
			where, following Eq.~\eqref{eq:HolevoDeriv}, we define $\chi^{(2)}_X(\rho^{AX}) \coloneqq \frac{1}{2}\partial_\mu^2 \chi_X^{\vphantom{(2)}}(\rho^{AB}, F_a^A(\mu))\big|_{\mu = 0}$.
		\end{lemma}
		We can use the above result twice with $X$ replaced by $B$ and $E$, and employing the same steps as before we find that the differences in $\chi_B^{(2)}$ and $\chi_E^{(2)}$ between the exact target state $\rho^{ABC}$ and the $\epsilon$-approximate state $\sigma^{ABC}$ is at most $6\epsilon$ each. Combined with the triangle inequality, this gives us Eq.~\eqref{eq:KeyContDeriv}. We now prove Lemma \ref{lem:HolevoContDeriv}.\\

		Without loss of generality we assume $\chi_X^{(2)}(\rho^{AX}) \geq \chi_X^{(2)}(\sigma^{AX})$. Let us define a path of density matrices $\rho^{AX}_t = (1-t)\rho^{AX} +t \sigma^{AX}$, and view $f(t) \coloneqq \chi_X^{(2)}(\rho^{AX}_t)$ as a scalar function of $t$. By the mean value theorem, there exists a $t \in [0,1]$ such that $\frac{\dif f(t)}{\dif t} = \chi_X^{(2)}(\rho^{AX}) - \chi_X^{(2)}(\sigma^{AX})$.
		Now, by Eq.~\eqref{eq:Chi2General}, we can rewrite our expression for $\chi_X^{(2)}$ as follows
		\begin{align}
			f(t) \coloneqq \frac{1}{2}\chi_X^{(2)}(\rho^{AX}_t) = \frac{1}{2}\left(\Tr\big[\xi_{t}^X\mathcal{T}_{\rho^X_t}[\xi_{t}^X]\big]-\Tr\big[\rho^A_t O^A\big]^2\right),
			\label{eq:Chi2T}
		\end{align}
		where $\xi_{t} \coloneqq \Tr_A[O^A \rho^{AX}_t]$ and the state-dependent map $\mathcal{T}_\sigma$ is defined in Eq.~\eqref{eq:TDef}. 
		Differentiating the expression \eqref{eq:Chi2T} we find
		\begin{align}
			\frac{\dif f(t)}{\dif t} =  \Tr\big[(\xi_{1}^X-\xi_{0}^X)\mathcal{T}_{\rho^X_t}[\xi_{t}^X] + \Tr\big[(\rho^X_1 - \rho^X_0)\mathcal{R}_{\rho^X_t}[\xi_{t}^X]\big] -\Tr[\rho^A_t O^A]\Tr[(\rho^A_1 - \rho^A_0)O^A]
		\end{align}
		where, again using the notation of Ref.~\cite{Lieb1973}, we define the $\rho$-dependent quadratic map
		\begin{align}
			\mathcal{R}_\rho[X] \coloneqq \int_0^\infty \dif z\, (\rho+zI)^{-1}X(\rho+zI)^{-1}X(\rho+zI)^{-1}
			\label{eq:RMapDef}
		\end{align}
		Below, we prove the following inequalities
		\begin{align}
			\|\xi_{1}^X - \xi_{0}^X\|_1 &\leq \|\rho^{AX} - \sigma^{AX}\|_1, \label{eq:Ineq1}\\
			\|\xi_{t}^X\|_1 &\leq 1, \label{eq:Ineq2}\\
			\|\mathcal{T}_{\rho^X_t}[\xi_{t}^X]\|_\infty &\leq 1, \label{eq:Ineq3}\\
			\|\mathcal{R}_{\rho^X_t}[\xi_{t}^X]\|_\infty &\leq 1. \label{eq:Ineq4}
		\end{align}
		Using H{\"o}lder's inequality $\Tr[XY]\leq \|X\|_1\|Y\|_\infty$, the fact that $\|O^A\|_\infty \leq 1$, and the data processing inequality for the trace distance $\|\sigma^X - \rho^X\|_1 \leq \|\sigma^{AX} - \rho^{AX}\|_1$, we can bound each term separately and find that
		\begin{align}
			\frac{\dif f(t)}{\dif t} \leq 3\|\rho^{AX} - \sigma^{AX}\|_1
		\end{align}
		which proves the lemma. We finally prove the inequalities (\ref{eq:Ineq1}--\ref{eq:Ineq4}). The first two can be derived from the dual characterization of the trace norm $\|R\|_1 = \sup_{T : \|T\|_\infty \leq 1} \Tr[RT]$. From the definition of $\xi_{t}^X$ and using $\|O^A\|_\infty \leq 1$, we have
		\begin{align}
			\|\xi_{1}^X - \xi_{0}^X\|_1 &= \sup_{T^X : \|T^X\|_\infty \leq 1} \Tr[( O^A \otimes T^X )(\sigma^{AX}-\rho^{AX})] \nonumber\\ &\leq   \sup_{T^{AX} : \|T^{AX}\|_\infty \leq 1} \Tr[T^{AX}(\sigma^{AX}-\rho^{AX})] = \|\rho^{AX} - \sigma^{AX}\|_1,
		\end{align}
		which is Eq.~\eqref{eq:Ineq1}. Eq.~\eqref{eq:Ineq2} follows in the same way using $\rho^{AX}$ in place of $(\sigma^{AX} - \rho^{AX})$.
		To prove the remaining two inequalities, we first show that
		\begin{align}
			-\rho^X_t \preceq \xi_{t}^X \preceq \rho^X_t,
			\label{eq:OperatorOrder}
		\end{align}
		where for two Hermitian matrices $X, Y$, the notation $X \preceq Y$ implies that that $(Y-X)$ is positive semi-definite. Let $\{\ket{j}_A\}$ be an eigenbasis of $O^A$ with corresponding eigenvectors $\lambda_j$. Then, for an arbitrary state $\ket{\phi} \in \mathcal{H}^X$,
		\begin{align}
			\braket{\phi|(\rho^X_t \pm \xi_{t}^X) |\phi} &=  \sum_j \Braket{\phi\otimes j|\big(I^X\otimes (I^A \pm O^A)\big) \rho^{AX}|\phi \otimes j}  \nonumber\\
			&=  \sum_j(1 \pm \lambda_j) \braket{\phi\otimes j|\rho^{AX}|\phi \otimes j}
			\geq 0
		\end{align}
		where in the last step we use $\rho^{AX} \succeq 0$, and also $-1 \leq \lambda_j \leq 1$, on account of the fact that $\|O^A\|_\infty \leq 1$. This proves Eq.~\eqref{eq:OperatorOrder}. Now, for any $\rho$, the map $\mathcal{T}_{\rho}$ is completely positive, as can be seen by interpreting its definition \eqref{eq:TDef} as a Kraus representation \cite{Lieb1973}. One can also verify that $\mathcal{T}_{\rho}[\rho] = I$. Therefore, by applying $\mathcal{T}_{\rho_t^X}$ to each part of \eqref{eq:OperatorOrder}, we get
		\begin{align}
			-I \preceq \mathcal{T}_{\rho^X_t}[\xi^X_t] \preceq I.
		\end{align}
		This in turn implies Eq.~\eqref{eq:Ineq3}. For the last inequality \eqref{eq:Ineq4}, we again start from Eq.~\eqref{eq:OperatorOrder} and conjugate on the left and right by $(\rho^X_t)^{-1/2}$ to get $-I \preceq (\rho^X_t)^{-1/2}\xi_{t}^X(\rho^X_t)^{-1/2} \preceq I$. Taking the square of this operator, we get
		\begin{align}0 \preceq (\rho^X_t)^{-1/2}\xi_{t}^X(\rho^X_t)^{-1} \xi_{t}^X (\rho^X_t)^{-1/2} \preceq I
		\end{align}
		Since $(\rho^X_t + zI)^{-1} \preceq (\rho^X_t)^{-1}$ for any $z \geq 0$, we can infer that $0 \preceq \xi_{t}^X (\rho^X_t + zI)^{-1}\xi_{t}^X \preceq \rho^X_t$, which can be substituted into the definition \eqref{eq:RMapDef} to get
		\begin{align}
			0 \preceq \mathcal{R}_{\rho^X_t}[\xi_{t}^X] \preceq \int_0^\infty \dif z\, (\rho^X_t+zI)^{-1} \rho^X_t (\rho^X_t+zI)^{-1} = I
		\end{align}
		which gives us Eq.~\eqref{eq:Ineq4}. \hfill $\square$

		\section{Computing the Holevo information in  critical thermal states}
		
		Here we evaluate the Holevo information for regions $B$ and $E$ for a weak measurement of an operator $O^A$ of strength $\mu$ for the Gibbs state of a critical Hamiltonian $H$. In particular, we look to compute the second order in $\mu$ contributions $\chi_{B,E}^{(2)}$ [Eq.~\eqref{eq:HolevoDeriv}]. We work in the regime of long wavelengths and low temperatures, such that physical quantities can be described by a ($1+1$)-diensional field theory with Lagrangian $\mathcal{L}$ that is conformally invariant. The observable $O^A$ in the microscopic theory is identified with a primary field $O(x)$ at the origin $x = 0$, whose scaling dimension is $\Delta$. The region $B$ will be a finite strip $[x_1,x_2]$, with $0 < x_1 < x_2$. In describing field-theoretic operators, we will work in units of space and time where the velocity of the CFT is unity, and we will use complex coordinates $z = x + \iu \tau$, where $\tau$ is the Euclidean time, such that conformal transformations $z \mapsto w$ correspond to holomorphic functions $w(z)$.
		
		\subsection{Computing $\chi_E^{(2)}$}
		
		Thanks to our expression \eqref{eq:ChiEThermal}, $\chi_E^{(2)}$ can be evaluated just from the dynamical autocorrelation function $C_\beta(O^A, t)$ at temperature $\beta^{-1}$. This is a standard exercise in CFT (see e.g.~Ref.~\cite{Blote1986}), but we go through the derivation here as several techniques will be used again later. We consider the imaginary-time autocorrelation function of the microscopic operator $O^A$, which can be analytically continued to real time at the end, and identify this with the corresponding correlator in the field theory
		\begin{align}
			C_\beta(O,\tau) = \frac{\Tr[O(\iu \tau)O(0) e^{-\beta H}]}{\Tr[e^{-\beta H}]} = \braket{O(\iu \tau)O(0)}_{\mathcal{L}, \mathcal{C}^{(\beta)}}
		\end{align}
		Here, the notation $\braket{\,\cdot\,}_{\mathcal{L}, \mathcal{C}^{(\beta)}}$ denotes a correlation function of the field theory with Lagrangian $\mathcal{L}$ on the manifold $\mathcal{C}^{(\beta)}$, which here is a cylinder of infinite spatial extent and radius $\beta$ in the time direction. The conformal transformation
		\begin{align}
			w \mapsto z(w) = e^{2\pi w/\beta}
			\label{eq:ConformalBeta}
		\end{align}
		maps the cylinder $\mathcal{C}^{(\beta)}$ to the complex plane $\mathbbm{C}$, and since $O(w)$ is a primary field we can use the defining relation
		\begin{align}
			\braket{O(w_1) O(w_2)}_{\mathcal{L}, \mathcal{C}^{(\beta)}} = \left| \frac{\partial w_1}{\partial z_1}\right|^{-\Delta}\left| \frac{\partial w_2}{\partial z_2}\right|^{-\Delta}\braket{O(z_1) O(z_2)}_{\mathcal{L}, \mathbb{C}}.
		\end{align}
		Since both the Lagrangian $\mathcal{L}$ and the manifold $\mathbbm{C}$ are invariant under translations and scaling transformations, we have
		\begin{align}
			\braket{O(z_1) O(z_2)}_{\mathcal{L}, \mathbb{C}} = \frac{\kappa_O}{|z_1 - z_2|^{2\Delta}}
			\label{eq:CFTTwoPoint}
		\end{align}
		where $\kappa_O$ is a non-universal constant. Using $\partial w/\partial z = \frac{\beta}{2\pi} e^{-2\pi w/\beta}$, this gives
		\begin{align}
			C_\beta(O,\tau) = \kappa_O\left| \frac{2\pi}{\beta} \frac{e^{\pi \iu \tau/\beta}}{e^{2\pi \iu \tau/\beta} - 1} \right|^{2\Delta} = \kappa_O(\pi T)^{2\Delta} |\sin(\pi \tau T)|^{-2\Delta}
		\end{align}
		Analytically continuing back to real time, we find
		\begin{align}
			C_\beta(O,t) = \kappa_O (\pi T)^{2\Delta}|\sinh(\pi t T)|^{-2\Delta}
		\end{align}
		The above expression can be written as $\kappa_O(\pi T)^{2\Delta}f_\Delta(tT)$ for a dimensionless function $f_\Delta(x) \coloneqq |\sinh(\pi x)|^{-2\Delta}$ that only depends on $\Delta$, thus the frequency-space spectral function is $C_\beta(O, \omega) = \kappa_O (\pi T)^{2\Delta} \frac{\tilde{f}_\Delta(\omega/T)}{T}$, where $\tilde{f}_\Delta(k)$ is the Fourier transform of $f_\Delta(x)$. Combining this with \eqref{eq:ChiEThermal} we can infer
		\begin{align}
			\chi^{(2)}_E &=  \frac{1}{4\pi}\int_{-\infty}^\infty \dif \omega\, C_\beta(O,\omega) \frac{\beta \omega}{e^{\beta \omega} - 1}  =   \kappa_O \alpha_\Delta (\pi T)^{2\Delta} & \text{where }\alpha_\Delta \coloneqq \frac{1}{4\pi} \int_{-\infty}^{\infty} \dif u \frac{u}{e^u - 1} \tilde{f}_\Delta(u)
			\label{eq:ChiEThermal1}
		\end{align}
		Note that the universal constant $\alpha_\Delta $ depends only on  the scaling dimension $\Delta$ (we will recover its exact value later on).
		
		\subsection{Computing $\chi_B^{(2)}$}
		
		Here, we will compute $\chi_B^{(2)}$ for an arbitrary connected interval $B = [x_1, x_2]$ with $x_2 > x_1$. When computing von Neumann entropies of reduced states of a conformal field theory (and associated linear combinations thereof), a standard approach is to use the replica trick (see Ref.~\cite{Calabrese2009}), where one first computes the $n$th R{\'e}nyi entropy $S^{(n)}(\rho) = (1-n)^{-1} \ln \Tr[\rho^n]$ for arbitrary integer $n$, takes an analytic continuation to non-integer values of $n$, and finally takes the limit $ \lim_{n\rightarrow 1} S^{(n)}(\rho) = S(\rho)$. We therefore define the $n$-R{\'e}nyi Holevo information for region $B$
		\begin{align}
			\chi_{B, n} \coloneqq  -(n-1)^{-1}\left(\sum_a  \Big[\ln \Tr[(\Tr_{A}[\rho^{AB} F_a^A])^n] - \ln(\Tr[\rho^A F^A_a]^n)\Big] - \Tr[(\rho^B)^n]\right)
		\end{align}
		which, as one can verify, reproduces the usual Holevo information in the appropriate limit $n \rightarrow 1$. Recalling that $F_a^A = \frac{1}{2}(I^A + (-1)^a \mu O^A)$, we take the contributions at second order in $\mu$, giving
		\begin{align}
			\chi_{B, n}^{(2)} \coloneqq \frac{1}{2} \left. \frac{\partial^2 \chi_{B,n}}{\partial \mu^2}\right|_{\mu = 0} = \sum_{\substack{j,k=1 \\ j \neq k}}^n \frac{\Tr[(O^{A_j}O^{A_k} \otimes \pi^{B_1\cdots B_n}) \cdot (\rho^{AB})^{\otimes n}]}{\Tr[(I^{A_1 \cdots A_n} \otimes\pi^{B_1 \cdots B_n})\cdot (\rho^{AB})^{\otimes n}]},
		\end{align}
		where $\pi^{B_1\cdots B_n}$ is a cyclic permutation operator. The trace above involves operators that act on $n$ replicas of the system, and $O^{A_j}$ is shorthand for the operator that acts as $O^A$ on the $j$th replica, and identity on all others. We start by considering zero temperature $\beta = \infty$ first. The ratio inside the sum can then be written as a correlation function, by identifying the denominator as a partition function which normalises the expectation value in the numerator,
		\begin{align}
			\frac{\Tr[(O^{A_j}O^{A_k} \otimes \pi^{B_1\cdots B_n}) \cdot (\rho^{AB})^{\otimes n}]}{\Tr[(I^{A_1 \cdots A_n} \otimes\pi^{B_1 \cdots B_n})\cdot (\rho^{AB})^{\otimes n}]} = \braket{O(w_j) O(w_k)}_{\mathcal{L}, \mathcal{M}_{n,x_1,x_2}} \label{eq:RiemannCorr}
		\end{align}
		Here, $\braket{\,\cdot\,}_{\mathcal{L}, \mathcal{M}_{n,x_1,x_2}}$ denotes an expectation value in a replica field theory defined on a $n$-sheeted Riemannian surface $\mathcal{M}_{n,x_1,x_2}$, where adjascent sheets are connected to each other in the interval $B = [x_1,x_2]$, with Lagrangian density $\mathcal{L}$ (See Section 2 of Ref.~\cite{Cardy2007}). Here, $w_j$ is the origin ($x = 0$) on the $j$th sheet of $\mathcal{M}_{n,x_1,x_2}$ corresponding to the location of the measurment operator $O^A$. Now, the conformal transformation
		\begin{align}
			w \mapsto z(w) = \left(\frac{w-x_1}{w-x_2}\right)^{1/n}
			\label{eq:ConfZeroTemp}
		\end{align}
		maps $\mathcal{M}_{n,x_1,x_2}$ onto $\mathbb{C}$, where here we choose the branch for the $n$th root according to which of the $n$ sheets $w$ belongs. Because the Lagrangian of the field theory $\mathcal{L}$ is conformally invariant and $O(w)$ is a primary field, the correlation function in question \eqref{eq:RiemannCorr} can be related to an analogous correlator of the same theory defined on the complex plane
		\begin{align}
			\braket{A(w_j) A(w_k)}_{\mathcal{L}, \mathcal{M}_{n,x_1,x_2}} = \left| \frac{\partial w_j}{\partial z_j}\right|^{-\Delta}\left| \frac{\partial w_k}{\partial z_k}\right|^{-\Delta}\braket{A(z_j) A(z_k)}_{\mathcal{L}, \mathbb{C}}.
		\end{align}
		Since $w_{j,k}$ correspond to the origins ($x = \tau = 0$) of sheets $j,k$, we identify the corresponding coordinates after the transformation \eqref{eq:ConfZeroTemp} as $z_j = e^{2\pi \iu j/n}(x_1/x_2)^{1/n}$. We also have
		\begin{align}
			\frac{\partial z(w)}{\partial w} = \frac{z}{n}\frac{(x_1-x_2)}{(w-x_1)(w-x_2)}
		\end{align}
		and from the scaling behaviour of the correlation function of the CFT on the complex plane $\mathbb{C}$ [Eq.~\eqref{eq:CFTTwoPoint}], we can infer
		\begin{align}
			\left.\frac{\Tr[(O^{A_j}O^{A_k} \otimes \pi^{B_1\cdots B_n}) \cdot (\rho^{AB})^{\otimes n}]}{\Tr[(I^{A_1 \cdots A_n} \otimes\pi^{B_1 \cdots B_n})\cdot (\rho^{AB})^{\otimes n}]}\right|_{\beta = \infty} = \kappa_O\left(\frac{1}{2n} \cdot \frac{|x_2-x_1|}{|x_1||x_2|}\right)^{2\Delta} \left|\sin\left(\frac{\pi (j-k)}{n}\right)\right|^{-2\Delta}
			\label{eq:ReplicaCorrelatorZero}
		\end{align}
		Taking the required sum over $j,k$ and then taking the limit $n\rightarrow 1$, we obtain the Holevo information to second order in $\mu$
		\begin{align}
			\chi^{(2)}_B(\beta = \infty) =&\, \kappa_O h(\Delta) \left(\frac{|x_2-x_1|}{2 |x_1||x_2|}\right)^{2\Delta}  
			\\
			\text{where }h(\Delta) \coloneqq& \lim_{n\rightarrow 1}\frac{n}{1-n} \sum_{m=1}^{n-1}\sin(m\pi/n)^{-2\Delta} = \frac{\sqrt{\pi} \Gamma(\Delta+1)}{2\Gamma(\Delta + 3/2)}
		\end{align}
		The necessary analytic continuation of the sum in the above expression for $h(\Delta)$ is treated in Section 6 of Ref.~\cite{Calabrese2011}.
		
		At finite temperature, the relevant correlation function in place of \eqref{eq:RiemannCorr} instead involves a $n$-sheet Riemann surface $\mathcal{M}_{n,x_1,x_2}^{(\beta)}$ where each replica is itself a cylinder of circumference $\beta$ (as opposed to the infinite complex plane), and adjascent sheets are joined along the branch cut $[x_1,x_2]$. By analogy to Eq.~\eqref{eq:ConformalBeta}, coordinates $u$ on this new manifold can be conformally mapped onto the manifold we considered before $\mathcal{M}_{n,x_1,x_2}^{(\beta = \infty)}$ via the transformation $u \mapsto w(u) = e^{2\pi u/\beta}$. By concatenating this map with \eqref{eq:ConfZeroTemp}, we can generalise \eqref{eq:ReplicaCorrelatorZero} to finite temperatures via
		\begin{align}
			\left.\frac{\Tr[(O^{A_j}O^{A_k} \otimes \pi^{B_1\cdots B_n}) \cdot (\rho^{AB})^{\otimes n}]}{\Tr[(I^{A_1 \cdots A_n} \otimes\pi^{B_1 \cdots B_n})\cdot (\rho^{AB})^{\otimes n}]}\right|_{\beta} = \kappa_O \left(\frac{\pi T}{n}\right)^{2\Delta} \left|\frac{\sinh(\pi T(x_2-x_1))}{\sinh(\pi Tx_1)\sinh(\pi T x_2)}\right|^{2\Delta} \left|\sin\left(\frac{\pi (j-k)}{n}\right)\right|^{-2\Delta}
		\end{align}
		which gives us our general result
		\begin{align}
			\chi^{(2)}_B = \kappa_O h(\Delta) (\pi T)^{2\Delta} \left|\frac{\sinh(\pi T(x_2-x_1))}{\sinh(\pi Tx_1)\sinh(\pi T x_2)}\right|^{2\Delta} 
		\end{align}
		Taking the limits $x_2 \rightarrow +\infty$, $x_1 \rightarrow -\infty$, we expect to recover our previous expression \eqref{eq:ChiEThermal1} for $\chi^{(2)}_E$. This is indeed the case, and we can thereby identify the two universal constants as $\alpha_\Delta = 2^{2\Delta}h(\Delta)$. We consider a geometry for $B$ with $x_1 = x_{AB} > 0$ and take $x_2 \rightarrow \infty$, in which case the private information appearing on the left hand side of \eqref{eq:KeyBound} becomes
		\begin{align}
			K^{(2)} = \chi^{(2)}_B - \chi^{(2)}_E = \kappa_O h(\Delta) (2\pi T)^{2\Delta}\left[\left(\frac{1}{e^{2\pi T x_{AB}} - 1}\right)^{2\Delta} - 1\right]
		\end{align}
		For exact state preparation, the necessary bound is $K^{(2)} > 0$, which immediately gives us a complexity lower bound of $d_{\rm min}(\rho_\beta) > \beta \ln(2)/4\pi$. For $\epsilon$-approximate state preparation, the criterion from Theorem \ref{thm:Approx} is $K^{(2)} > 12\epsilon$, and so the lower bound on circuit depth can be read off as
		\begin{align}
			d_{\rm min}(\rho_\beta, \epsilon) \geq \frac{\beta}{4\pi} \ln \left(1 + \frac{1}{(1 + c \beta^{2\Delta} \epsilon)^{1/2\Delta}}\right)
		\end{align}
		where $c = \frac{12}{(2\pi)^{1/2\Delta }\kappa_Oh(\Delta)}$ is a non-universal constant. Since we could have chosen any primary operator $O(x)$ from the beginning, we are free to take the scaling dimension $\Delta$ to be the smallest in the theory, $\Delta_{\rm min}$, which gives the best dependence on $\epsilon$, and we thereby obtain the formula \eqref{eq:CFTLowerBound} quoted in the main text.
		
	\end{onecolumngrid}

\end{document}